\documentclass[12pt,fleqn]{book} 
\usepackage[top=1.0in,bottom=1.0in,left=1.0in,right=1.0in,headsep=10pt,letterpaper]{geometry} 
\usepackage{xcolor} 
\definecolor{ocre}{RGB}{52,177,201} 

\usepackage{avant} 
\usepackage{mathptmx} 

\usepackage{microtype} 
\usepackage[utf8]{inputenc} 
\usepackage[T1]{fontenc} 

\usepackage{titlesec} 

\usepackage{graphicx} 
\graphicspath{{Pictures/}{figures/cosmo/}} 

\usepackage{lipsum} 

\usepackage{tikz} 

\usepackage[english]{babel} 

\usepackage{enumitem} 
\setlist{nolistsep} 

\usepackage{booktabs} 

\usepackage{eso-pic} 

\usepackage{titletoc} 

\contentsmargin{0cm} 
\titlecontents{chapter}[1.25cm] 
{\addvspace{15pt}\large\sffamily\bfseries} 
{\color{ocre!60}\contentslabel[\Large\thecontentslabel]{1.25cm}\color{ocre}} 
{}
{\color{ocre!60}\normalsize\sffamily\bfseries\;\titlerule*[.5pc]{.}\;\thecontentspage} 
\titlecontents{section}[1.25cm] 
{\addvspace{5pt}\sffamily\bfseries} 
{\contentslabel[\thecontentslabel]{1.25cm}} 
{}
{\sffamily\hfill\color{black}\thecontentspage} 
[]
\titlecontents{subsection}[1.25cm] 
{\addvspace{1pt}\sffamily\small} 
{\contentslabel[\thecontentslabel]{1.25cm}} 
{}
{\sffamily\;\titlerule*[.5pc]{.}\;\thecontentspage} 
[]

\titlecontents{lsection}[0em] 
{\footnotesize\sffamily} 
{}
{}
{}

\titlecontents{lsubsection}[.5em] 
{\normalfont\footnotesize\sffamily} 
{}
{}
{}

\usepackage{fancyhdr} 

\fancypagestyle{plain}{\fancyhead{}} 

\makeatletter
\renewcommand{\cleardoublepage}{
\clearpage\ifodd\c@page\else
\hbox{}
\vspace*{\fill}
\thispagestyle{empty}
\newpage
\fi}

\usepackage{amsmath,amsfonts,amssymb,amsthm} 

\newtheoremstyle{ocrenumbox}
{0pt}
{0pt}
{\normalfont}
{}
{\small\bf\sffamily\color{ocre}}
{\;}
{0.25em}
{\small\sffamily\color{ocre}\thmname{#1}\nobreakspace\thmnumber{\@ifnotempty{#1}{}\@upn{#2}}
\thmnote{\nobreakspace\the\thm@notefont\sffamily\bfseries\color{black}---\nobreakspace#3.}} 

\newtheoremstyle{blacknumex}
{5pt}
{5pt}
{\normalfont}
{} 
{\small\bf\sffamily}
{\;}
{0.25em}
{\small\sffamily{\tiny\ensuremath{\blacksquare}}\nobreakspace\thmname{#1}\nobreakspace\thmnumber{\@ifnotempty{#1}{}\@upn{#2}}
\thmnote{\nobreakspace\the\thm@notefont\sffamily\bfseries---\nobreakspace#3.}}

\newtheoremstyle{blacknumbox} 
{0pt}
{0pt}
{\normalfont}
{}
{\small\bf\sffamily}
{\;}
{0.25em}
{\small\sffamily\thmname{#1}\nobreakspace\thmnumber{\@ifnotempty{#1}{}\@upn{#2}}
\thmnote{\nobreakspace\the\thm@notefont\sffamily\bfseries---\nobreakspace#3.}}

\newtheoremstyle{ocrenum}
{5pt}
{5pt}
{\normalfont}
{}
{\small\bf\sffamily\color{ocre}}
{\;}
{0.25em}
{\small\sffamily\color{ocre}\thmname{#1}\nobreakspace\thmnumber{\@ifnotempty{#1}{}\@upn{#2}}
\thmnote{\nobreakspace\the\thm@notefont\sffamily\bfseries\color{black}---\nobreakspace#3.}} 
\makeatother

\newcounter{dummy}
\numberwithin{dummy}{section}
\theoremstyle{ocrenumbox}
\newtheorem{theoremeT}[dummy]{Theorem}

\newtheorem{exerciseT}{Exercise}[chapter]
\theoremstyle{blacknumex}
\newtheorem{exampleT}{Example}[chapter]
\theoremstyle{blacknumbox}

\newtheorem{definitionT}{Definition}[section]
\newtheorem{corollaryT}[dummy]{Corollary}
\theoremstyle{ocrenum}

\RequirePackage[framemethod=default]{mdframed} 

\newmdenv[skipabove=7pt,
skipbelow=7pt,
backgroundcolor=black!5,
linecolor=ocre,
innerleftmargin=5pt,
innerrightmargin=5pt,
innertopmargin=5pt,
leftmargin=0cm,
rightmargin=0cm,
innerbottommargin=5pt]{tBox}

\newmdenv[skipabove=7pt,
skipbelow=7pt,
rightline=false,
leftline=true,
topline=false,
bottomline=false,
backgroundcolor=ocre!10,
linecolor=ocre,
innerleftmargin=5pt,
innerrightmargin=5pt,
innertopmargin=5pt,
innerbottommargin=5pt,
leftmargin=0cm,
rightmargin=0cm,
linewidth=4pt]{eBox}	

\newmdenv[skipabove=7pt,
skipbelow=7pt,
rightline=false,
leftline=true,
topline=false,
bottomline=false,
linecolor=ocre,
innerleftmargin=5pt,
innerrightmargin=5pt,
innertopmargin=0pt,
leftmargin=0cm,
rightmargin=0cm,
linewidth=4pt,
innerbottommargin=0pt]{dBox}	

\newmdenv[skipabove=7pt,
skipbelow=7pt,
rightline=false,
leftline=true,
topline=false,
bottomline=false,
linecolor=gray,
backgroundcolor=black!5,
innerleftmargin=5pt,
innerrightmargin=5pt,
innertopmargin=5pt,
leftmargin=0cm,
rightmargin=0cm,
linewidth=4pt,
innerbottommargin=5pt]{cBox}

\makeatletter
\renewcommand{\@seccntformat}[1]{\llap{\textcolor{ocre}{\csname the#1\endcsname}\hspace{1em}}}
\renewcommand{\section}{\@startsection{section}{1}{\z@}
{-2ex \@plus -1ex \@minus -.2ex}
{1ex \@plus.1ex }
{\normalfont\large\sffamily\bfseries}}
\renewcommand{\subsection}{\@startsection {subsection}{2}{\z@}
{-2ex \@plus -0.1ex \@minus -.2ex}
{0.5ex \@plus.2ex }
{\normalfont\sffamily\bfseries}}
\renewcommand{\subsubsection}{\@startsection {subsubsection}{3}{\z@}
{-2ex \@plus -0.1ex \@minus -.2ex}
{.2ex \@plus.2ex }
{\normalfont\small\sffamily\bfseries}}
\renewcommand\paragraph{\@startsection{paragraph}{4}{\z@}
{-2ex \@plus-.2ex \@minus .2ex}
{.1ex}
{\normalfont\small\sffamily\bfseries}}

\usepackage{hyperref}
\hypersetup{hidelinks,backref=true,pagebackref=true,hyperindex=true,colorlinks=false,breaklinks=true,urlcolor= ocre,bookmarks=true,bookmarksopen=false,pdftitle={Title},pdfauthor={Author}}

\newcommand{\thechapterimage}{}
\newcommand{\chapterimage}[1]{\renewcommand{\thechapterimage}{#1}}

\def\thechapter{\arabic{chapter}}
\def\@makechapterhead#1{
\thispagestyle{empty}
{\centering \normalfont\sffamily
\ifnum \c@secnumdepth >\m@ne
\if@mainmatter
\startcontents
\begin{tikzpicture}[remember picture,overlay]
\node at (current page.north west)
{\begin{tikzpicture}[remember picture,overlay]
\node[anchor=north west,inner sep=0pt] at (0,0) {\includegraphics[width=\paperwidth]{\thechapterimage}};
\draw[anchor=west] (5cm,-9cm) node [rounded corners=20pt,fill=ocre!10!white,text opacity=1,draw=ocre,draw opacity=1,line width=1.5pt,fill opacity=.6,inner sep=12pt]{\huge\sffamily\bfseries\textcolor{black}{\thechapter. #1\strut\makebox[22cm]{}}};
\end{tikzpicture}};
\end{tikzpicture}}
\par\vspace*{230\p@}
\fi
\fi}

\def\@makeschapterhead#1{
\thispagestyle{empty}
{\centering \normalfont\sffamily
\ifnum \c@secnumdepth >\m@ne
\if@mainmatter
\begin{tikzpicture}[remember picture,overlay]
\node at (current page.north west)
{\begin{tikzpicture}[remember picture,overlay]
\node[anchor=north west,inner sep=0pt] at (0,0) {\includegraphics[width=\paperwidth]{\thechapterimage}};
\draw[anchor=west] (5cm,-6cm) node [rounded corners=20pt,fill=ocre!10!white,fill opacity=.6,inner sep=12pt,text opacity=1,draw=ocre,draw opacity=1,line width=1.5pt]{\LARGE\sffamily\bfseries\textcolor{black}{#1\strut\makebox[22cm]{}}};
\end{tikzpicture}};
\end{tikzpicture}}
\par\vspace*{130\p@}
\fi
\fi
}
\makeatother

\usepackage{graphicx}
\graphicspath{{./figures/}}
\usepackage{appendix}
\usepackage{latexsym,amsmath,amsfonts,amssymb,booktabs}
\usepackage[font=small]{caption}
\usepackage{slashed,upgreek,amscd,cancel,tensor,color}
\usepackage{adjustbox}
\usepackage[numbers,sort&compress,square]{natbib}
\usepackage{epsfig,latexsym}
\usepackage{url}
\numberwithin{equation}{section}
\usepackage{doi}
\usepackage{subcaption}
\usepackage{mathtools}
\usepackage{upgreek}
\usepackage{stfloats}
\usepackage{afterpage}
\usepackage{multirow}

\usepackage{aas_macros}
\usepackage{tcolorbox}
\usepackage[utf8]{inputenc}
\usepackage{wrapfig}

\begin{document}

\chapterimage{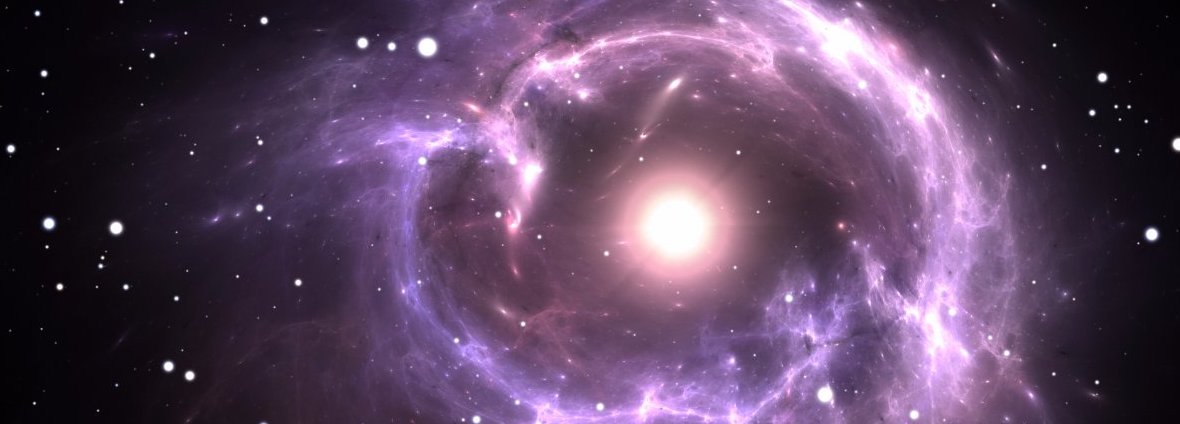} 
\chapter*{\Large The Yet-Unobserved GW Universe} 

  

\vspace{3cm}
\begin{center} 
\Large
\textbf{Astro2020 Science White Paper} \linebreak


THE YET-UNOBSERVED MULTI-MESSENGER GRAVITATIONAL-WAVE UNIVERSE 

\end{center} 

\normalsize

\vspace{-3cm}

\noindent \textbf{Thematic Areas:} 
\begin{itemize}
\item Formation and Evolution of Compact Objects 
\item Stars and Stellar Evolution 
\item Multi-Messenger Astronomy and Astrophysics
\end{itemize}


\vspace{0.75cm}

\noindent \textbf{Principal Author}: 

\vspace{-0.3cm}
 Name:	Vicky Kalogera
 \newline
 	
	\vspace{-0.8cm}					
Institution: Northwestern U.\ 
 \newline
 
 \vspace{-0.8cm}
Email: vicky@northwestern.edu
 \newline
 
 \vspace{-0.8cm}
Phone: +1-847-491-5669 
 \newline
 
 \noindent 
\textbf{Lead Co-authors:} Marrie-Anne Bizouard (CNRS, OCA), Adam Burrows (Princeton U.), Thomas Janka (MPA), Kei Kotake (Fukuoka U.), Bronson Messer (ORNL \& U. Tennessee),  Tony Mezzacappa (ORNL \& U. Tennessee), Bernhard Mueller (Monash U.), Ewald Mueller (MPA), Maria Alessandra Papa (AEI), Sanjay Reddy (U.\ Washington), Stephan Rosswog (Stockholms U.)

 \noindent 
 Click here for \href{https://docs.google.com/spreadsheets/d/1uNKEW77Fm-_nc21_3jSOX-P4ecRxsxA5UgDOD4bkG9Y/edit#gid=810591404}{\bf other co-authors and supporters}





\pagebreak


\chapterimage{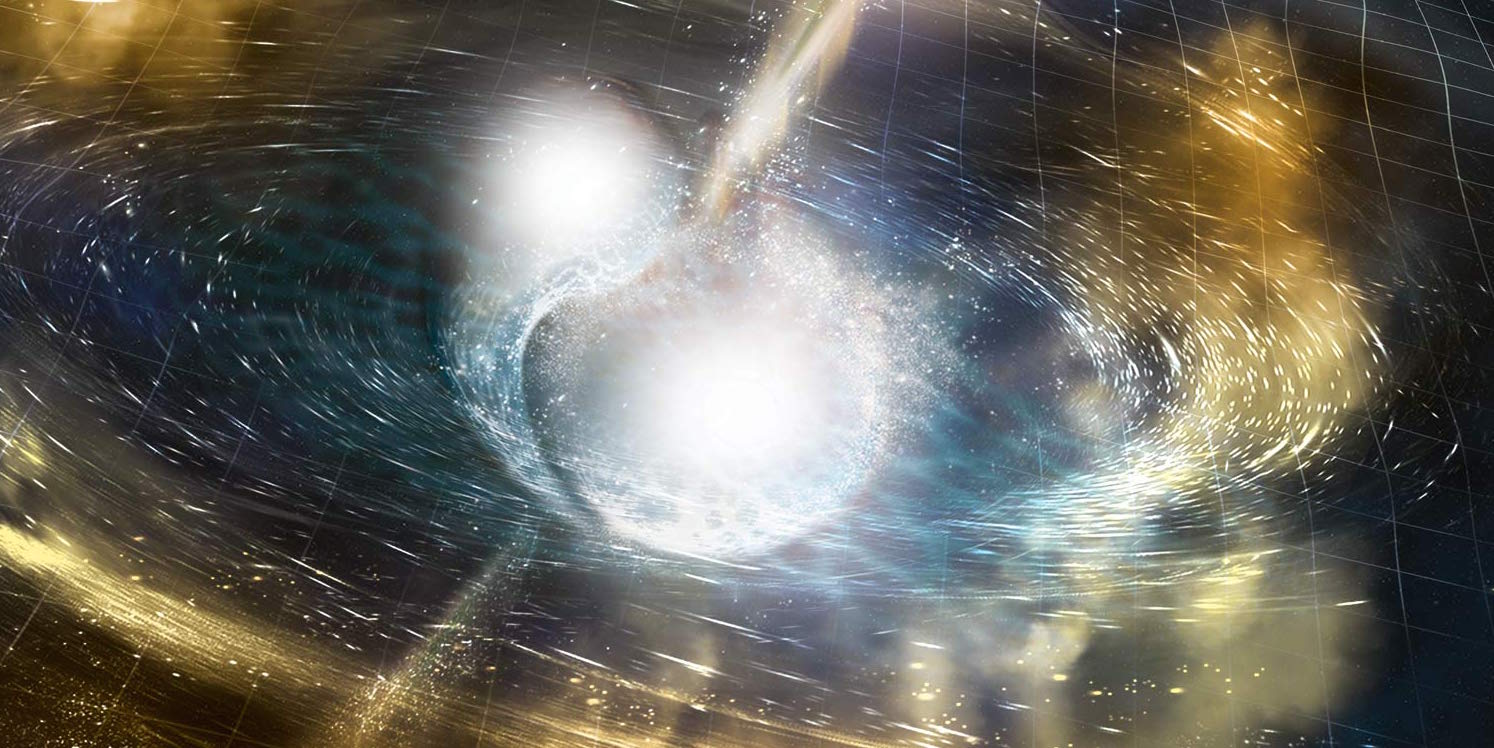} 

\section*{The Yet-unobserved Multi-messenger Gravitational-wave Universe}

Black holes (BHs) and neutron stars (NSs) have already been detected as chirping gravitational-wave (GW) sources \cite{Abbott:2016blz,LIGOScientific:2018mvr}, the latter also as a multi-messenger (MM) source with emission across the electromagnetic spectrum \cite{GBM:2017lvd}. However, BHs and NSs are predicted to be GW sources of burst or continuous-wave character, in isolation or in binary systems. 
These GW sources can also be MM sources, and combined MM observations will reveal richer details of the source astrophysics.  
Signal strengths are highly uncertain, but generally low, low enough that detection with the 2nd-generation detectors even at design sensitivity are far from guaranteed, if not impossible. Third-generation GW detectors will be necessary certainly for the reliable study of (i) bursts from the birth of compact objects when massive stars collapse as core-collapse supernovae (ccSN), (ii) bursts from magnetars or glitching radio pulsars, (iii) continuous GWs from NSs, isolated or in interacting binaries.

\begin{tcolorbox}[standard jigsaw,colframe=ocre,colback=ocre!10!white,opacityback=0.6,coltext=black]
Observations with next-generation GW detectors further enhanced with MM analyses of electromagnetic and possibly neutrino detections will allow us to probe new extreme astrophysics and answer key questions: 
\begin{itemize}[leftmargin=*]
\item {\bf Gravitational Waves from Core-Collapse Supernovae.} Which ccSN phases dominate the GW emission? Do the progenitors rotate and how fast? Does the event form a BH? 

\item {\bf Continuous GW Emission from Isolated or Accreting Neutron Stars.} What magnitude of deformations can NS crusts sustain and what are the implications for nuclear matter equation of state? Is the spin equilibrium of accreting NSs determined by GW emission and through what mechanism? 

\item {\bf Bursts from Magnetars and Other Pulsars.} Can GW detections help us probe the role of magnetic fields in transient emission from neutron stars and further constrain the equation of state of ultra dense matter?   

\end{itemize}
\end{tcolorbox}

 \section*{Core-Collapse Supernovae}

GWs are generated in core-collapse supernovae (ccSN) by time-dependent rotational flattening, proto-neutron-star (PNS) pulsations, 
non-axisymmetric bulk mass motions due to convection, non-radial accretion flows and instabilities, and other asymmetries  
associated with the effects of strong magnetic fields. 
The dominant GW emission occurs during the phase of neutrino-driven convection with the standing accretion shock instability (SASI)
and by  $\ell = 2$ f- and g-modes in the near-surface layers of the PNS \cite{murphy:09,yakunin:15,morozova}. At later times ($\sim$a few hundred ms), a single f-mode manifests itself in GW amplitude spectrograms as a narrow frequency band
whose location and width are determined by PNS properties \cite{morozova,MuJaMa13,kuroda:14,mueller:17}. The 3D models predict a well-defined structure in the
time--frequency domain, but the quadrupole amplitudes of these models are smaller than those of 2D models \cite{andresen:17,mueller:17}, as the downflows are decelerated before striking the PNS surface and also lack the necessary rapid time variability needed for resonant excitation of the f-mode. 

Each ccSN phase has a range of characteristic signatures in its GW signal that can provide diagnostic constraints on the evolution and physical parameters 
of the explosion and on the dynamics of the nascent PNS. {\bf Core Collapse and Bounce:} General-relativistic studies \cite{dimmelmeier:07,dimmelmeier:08,
richers:17} showed that the GW burst signal from core bounce has a 
generic shape \cite{ZwMu97} for a wide range of rotation rates and
rotation profiles; therefore, it is best for probing the bulk parameters of the collapsing iron core \cite{abdikamalov:14,fuller:15}. 
{\bf Neutrino-driven Turbulent Convection Outside and Inside the PNS:} Milliseconds after core bounce, prompt convection in the cavity between the PNS and standing shock produces a short period ($\sim$tens of ms) of GW activity 
peaking at $\sim 100$\,Hz. Several tens of ms post bounce, stochastic mass motions can lead to significant 
broadband emission ($10$--$500$\,Hz with a peak at about $100$--$200$\,Hz) \cite{kotake:09,mueller:12,andresen:17,MuRaBu04,murphy:09,yakunin:10,mueller:13,yakunin:15,morozova}. On the other hand, the typical properties of the inner PNS convection zone translate into a turnover timescale of ms, so the corresponding GW signal, also broadband, emerges in the range $500$\,Hz to a few kHz \cite{mueller:97,marek_08,yakunin:10,mueller:13,morozova,andresen:17}. {\bf PNS Oscillations:} Fundamental modes driven by gravity and pressure forces can generate GW emission. Most dominant are the quadrupolar g-modes \cite{torres:18} and the fundamental f-mode of the nascent PNS excited by the aspherical accretion of plumes of matter crashing onto it during the stalled accretion phase, as well as after explosion by the continuing fallback of matter 
\cite{murphy:08,cedraduran:13,morozova,torres:18}. The f- and g-modes can
also be excited by convection inside the PNS \cite{andresen:17}. 
The time--frequency trajectory of the dominant f-mode follows a well-defined path that is a direct function of the PNS mass, radius, and temperature \cite{mueller:12,torres:18}
and, hence, of the equation of state and of integrated neutrino losses. The excited frequencies are $\sim 200$--$500$ Hz in the early stage (within the first few hundred ms after
bounce)
and of $\sim500$--$2000$ Hz in the later stage (hundreds of ms to s after bounce)
\cite{morozova}. {\bf SASI:} This is an instability of the supernova shock itself. It exists in both 2D and 3D simulations, defined by a nonlinear,  sloshing mode in 2D, and by both sloshing and spiral modes in 3D \cite{BlMe07}. The SASI modulates the shock position on a time scale $\sim 50$ ms -- in turn modulating the accretion flow in the region below it -- i.e., the post-shock region, at frequencies $\simeq 100$--$250$\,Hz in both 2D and 3D \cite{MuRaBu04,murphy:09,yakunin:10,yakunin:15,kuroda:16,andresen:17,mueller:13}. 

Importantly, the onset of the neutrino emissions in ccSNe coincides (to within ms) with the onset of GW emission \cite{andresen:17,kuroda:16,yakunin:15,murphy:09,kotake:13}. The detection of neutrinos by Super-K/Hyper-K \cite{abe:2016}, DUNE \cite{ankowski:2016}, JUNO \cite{lu:2015}, IceCube \cite{abbasi:2011}, LVD \cite{Agafonova:2007hn}, Borexino \cite{Cadonati:2000kq}, KamLAND \cite{Tolich:2011zz}, and yet more sensitive neutrino detectors anticipated for the 2030's, will allow to optimally extract the GW signal \cite{kuroda:17}. Both signals are produced at the same interior locations resulting to not only in time coincidence, but are also correlated timescale modulations and polarization, which aids with signal extraction and interpretation. If the progenitor core is rotating, there are additional, distinctive modulation signatures \cite{tk18}. Joint MM analyses can not only enhance detectability but also more reliably probe physical processes, including those producing electromagnetic emission necessary for useful localization, for identification of host galaxies, progenitors, and potential progenitor binary companions. Synergistic observational strategies for optimizing MM campaigns for 
a future ccSN event have been articulated \cite{MM} in various situations. Third-generation GW detectors will be critical to extracting all the physics from observable ccSNe and will take advantage of EM/particle  detectors synergistically. 

The detection of GW signals from ccSNe will enable us to measure the progenitor mass, as it is one of the major determinants for parameters that affect the various signal components directly, such as the PNS mass or the violence of convective/SASI motions. Simulations show a qualitative trend in successful SN explosions towards stronger and longer GW emission for more massive progenitors \cite{yakunin:10,mueller:13,morozova}. The energy radiated in GWs is predicted to vary by several orders of magnitude. 
The core spin could be measured with GW and neutrino MM detections, as the GW frequency is twice the modulation frequency of the neutrino signal \cite{ott:12,yokozawa:15,kuroda:17,tk18}. Hot, nuclear matter EOS constraints are best obtained from the PNS modes and the SASI signals at $\sim 100$ to $250$\,Hz \cite{morozova,kuroda:16}. Time evolution of the GW frequency may allow us to probe the mass accretion history before and after shock revival, unless the process is purely stochastic. Bounce and explosion times are much harder to pin point, and if at all, require neutrino detections. A fundamental question is whether the ccSN explosion mechanism is neutrino- or MHD-driven but until MHD models are further developed, we are limited in our efforts  
If BH formation takes place in a rapidly spinning progenitor, it will be accompanied by an intense spike-like burst of GW emission at the point of relativistic collapse, followed by a fast ringdown as the newly formed BH settles down to a Kerr spacetime \cite{ott_11}. By contrast, BH formation during the first seconds after collapse in non-rotating or slowly rotating progenitors is likely to manifest itself only as an abrupt cutoff of GW emission after a long period of moderate-amplitude GW emission. 
Prior to BH formation, the characteristic frequencies of PNS oscillation modes in the spectrum will increase to several kHz \cite{cerda_13,pan:17}.
\begin{wrapfigure}{o}{0.5\textwidth}
\vspace{-0.7cm}
\begin{center}
\includegraphics[width=0.5\textwidth]{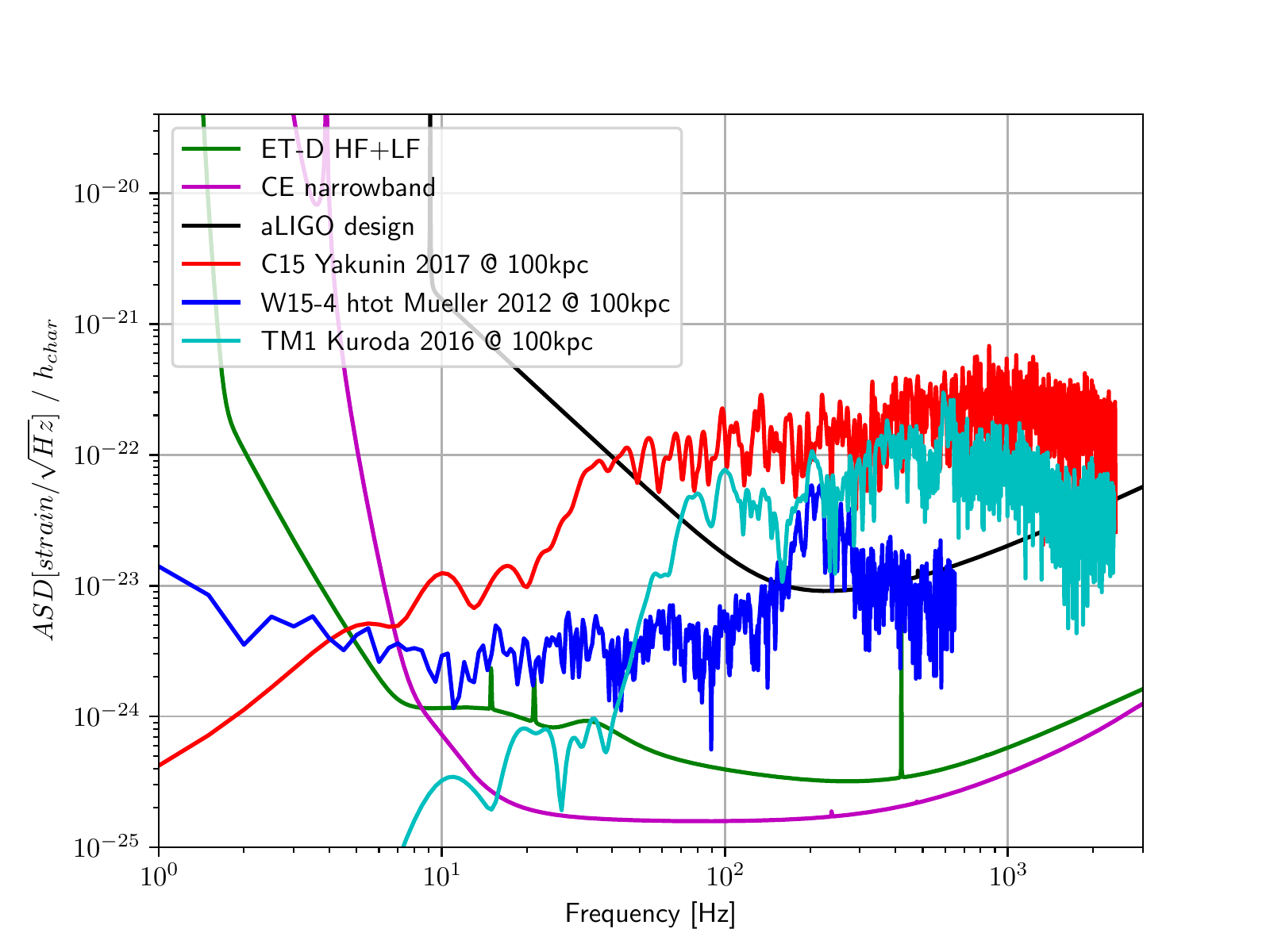}
\end{center}
\vspace{-0.7cm}
\caption{Characteristic strain vs.\ frequency of three typical 3D 
ccSN simulations: C15 \cite{yakunin:17}, W15-4 \cite{mueller:12}, 
and TM1 \cite{kuroda:16}. The Einstein Telescope 
in xylophone D (ET-D) configuration \cite{Hild:2010id}, the Cosmic Explorer (CE) \cite{Dwyer:2014fpa}, 
and the Advanced LIGO design \cite{Aasi:2013wya} also shown.}
\label{sn_fig1}
\end{wrapfigure}
\indent No ccSN GW signals have been detected so far. Even at design sensitivity, 2nd-generation detectors are not expected to reach outside our own Milky Way \cite{Gossan:2015xda}. Typical predicted 3D ccSN GW signals shown in Figure 1 for a source placed at $100$\,kpc have signal-to-noise ratios for Advanced LIGO at design in the range $0.5 - 6$, i.e., below reliable detectability levels. In contrast, they reach values in the range of $12$--$130$ for example designs of 3rd-generation detectors (Einstein Telescope and Cosmic Explorer). Therefore a range of $\sim 100$ kpc is a 
reasonable order-of-magnitude for the maximal detection distance of ccSN events. The goal of 3rd-generation detectors is not only to detect the signal (most likely aided by MM observations), not only to reconstruct the GW signal waveform and
the location of the source, but more importantly to determine with precision its intrinsic physical progenitor and explosion parameters (e.g., \cite{Heng:2008ww,powell:16}).

\section*{Sources of Continuous Gravitational Waves}

The emission of continuous high-frequency GWs at detectable amplitudes requires a time-varying mass quadrupole in a fast rotating compact object. They are expected whenever there is a sustained non-axisymmetric  distribution of matter in a rotating compact object \cite{2017MPLA...3230035R}. This can happen due to a variety of mechanisms. Most prominent examples include elastic stresses building up in the crust and giving rise to local deformations, deformations due to magnetic fields, which can occur in isolated NSs, and the growth of r-modes in accreting NSs (a fluid mode of oscillation for which the restoring force is the Coriolis force) \cite{1998ApJ...502..708A,1998ApJ...502..714F}. Whereas the amplitude of a GW signal depends on the details of the emission mechanism and on the source, the possible signal morphologies do not differ much. Typical continuous GWs are sinusoidal signals with a small spin-down or spin-up ($|\dot{f}|$ no larger than $10^{-7}$ Hz/s and most often smaller than $10^{-9}$ Hz/s) and a duration of at least a few weeks and most typically years.
As the loss rate of rotational energy caused by GW radiation is proportional
to the sixth power of the spin frequency, the most powerful sources must
possess rapid spin. Such large amounts of spin angular momentum can be a birth property of a newborn NS, or it may result from the recycling of an old
NS via accretion of matter and angular momentum from a companion star.

The timescale over which such a deformation can be sustained is crucial for detectability. In young NSs that are observable as radio pulsars and magnetars, a significant magnetic field is indeed present. Such NSs lose additional rotational energy to magneto-dipole radiation and magnetospheric currents, which reduces their GW luminosity considerably. From long-term monitoring of the braking index, it might be possible to distinguish between GW sources which spin down solely due to GW radiation from those which in addition spin down due to electromagnetic radiation.

The recycling of old NSs in interacting binaries may, in principle, allow for a population of NSs spinning 
at sub-ms spin periods -- and, equally important, supply the material needed to produce thermal or magnetic mountains. Whereas the fastest spinning ms pulsar is spinning at a frequency of 716~Hz, the mass-shedding frequency is still larger by a 
factor of 2. However, it is currently unclear whether magnetosphere-disk interactions can allow for the existence of sub-ms pulsars among the populations of low-mass X-ray binaries and radio ms pulsars. 

No continuous GW signal has so far been observed. The most current upper limits over the entire sky corresponds to a canonical NS at $10$ kpc, emitting GWs above 500 Hz (150 Hz) due to an ellipticity smaller than $10^{-5}$ ($10^{-7}$) \cite{Pisarski:2019vxw}; and upper limits on searches targetting known pulsars can be even stricter, with the limit for J0711$-$6830 at $1.2\times 10^{-8}$ \cite{Authors:2019ztc}. However, in order to detect strains factors of $\approx$ 100 lower than the current ones, 
new detectors, with a substantially lower noise floor are needed. Such sensitivity requires new, 3rd-generation detector facilities. 

The detection of continuous GWs from NSs in 3rd-generation detectors would be a fundamental breakthrough in our  attempts to peer into the ultra-dense interiors of NSs. It would provide clues about NS properties (isolated) or accretion and magnetosphere physics (binaries), their spin, thermal and magnetic field evolution,  the nature of cold dense matter, and phase transitions in QCD. Concurrent EM observations and input microphysics such as the transport coefficients and neutrino cooling rates will be essential to interpret these observations and harvest these fundamental insights.

{\bf Isolated Neutron Stars:} As first pointed out by Ruderman \cite{Ruderman:1969}, a solid NS crust can sustain (nonaxisymmetric) deformations. The maximum possible size of these deformations depends on the composition and structure of the crust. 
A fully general-relativistic calculation \cite{JohnsonMcDaniel:2012wg} (building on the Newtonian calculation in \cite{Ushomirsky:2000ax}) gives maximum fiducial ellipticities of $\sim 2\times 10^{-6}$ for the SLy EOS and its associated crustal model~\cite{Douchin:2001sv}; this EOS is consistent with LIGO observations of GW170817 (see, e.g., \cite{Abbott:2018exr}). Nevertheless, the fiducial ellipticities of $\sim 10^{-9}$ that are suggested to provide a floor on the spin-down of ms pulsars in \cite{Woan:2018tey} seem more likely to occur in a large population of stars than deformations near the theoretical maximum, particularly if one is only considering crustal deformations.
Magnetic fields significantly complicate the modeling of NS interiors.
However they play an important role in determining the spin evolution of NSs and possibly continuous GW emission. This is largely uncharted territory: developing fully relativistic MHD evolution models will be crucial to guide and interpret observations.

{\bf Accreting Neutron Stars in Binaries:} NSs in binary systems can also emit continuous GWs.
In fact, these systems might be more likely to present large deformations than their isolated siblings, as it may be possible for the NS surface magnetic
field to be compressed by infalling material,
such that a large quadrupolar ellipticity could be created
\cite{2005ApJ...623.1044M};
asymmetric heating of the interior due to accretion could lead to
sufficient thermal deformations that GWs are produced \cite{1998ApJ...501L..89B}; 
high-frequency oscillations during an X-ray burst
or outburst, could be due to a GW-emitting unstable r-mode
\cite{2014MNRAS.442.1786A,2014MNRAS.442.3037L}, e.g. as detected in
two accreting NSs \cite{2014ApJ...784...72S,2014ApJ...793L..38S}.
In fact, the emission of GWs could be the reason why we do not observe NSs 
spinning at their theoretical limit 
\cite{2003Natur.424...42C,2008AIPC.1068...67C,2017ApJ...850..106P}.
GWs from NSs in a binary could also uncover effects that may not
be studied in isolated NSs.
Long-term monitoring of the NS spin period, either via GWs only
or in combination with radio, X-ray, and/or gamma-ray observations, could
permit close tracking of both spin torques and orbital evolution. In these systems deciphering the physical mechanism responsible for GW emission will require MM observations.

\section*{Other GW Bursts from Magnetized Neutron Stars}

NSs can produce GW bursts, for example
via magnetar giant flares or pulsar glitches. If detected (e.g.,  after an
electromagnetic trigger), they can provide 
insights into the properties of high density matter. Current GW detectors have searched for such signals with no positive result.

Magnetars, highly magnetised NSs with magnetic fields exceeding
$10^{14}$\,G, are observed as anomalous X-Ray pulsars (AXP) or soft gamma-ray
repeaters (SGRs) \cite{kaspi17}. SGRs show recurrent X-ray activity that include
frequent short-duration bursts ($10^{36}-10^{43}$~erg~s$^{-1}$ with durations
of $\sim0.1$~s) and, in some cases, energetic giant flares \cite{turolla15}
($10^{44}-10^{47}$~erg~s$^{-1}$ within $0.1$~s with X-ray tails that can
extend to several $100$~s). Since they are thought to involve substantial
structural changes within the NSs and due to the large involved energy,
magnetars are potential GW sources, see \cite{lasky15,glampedakis17} for recent reviews.
They may, however, only be detectable if an energy corresponding to a significant fraction of the X-ray energy is channelled into GWs.

To date, three giant flares \cite{israel05,strohmayer05,strohmayer06} have been detected, and several bursts \cite{huppenkothen14a,huppenkothen14b}
have been observed that showed quasi-periodic oscillations (QPOs). A detection of magneto-elastic QPOs together with GW would provide incomparable information on the oscillation spectrum of NSs and thus allow to study their deep interior with unprecedented detail. For a NS at $10$~kpc with a magnetic field at the pole of $B_{\rm pole}\sim10^{15}$~G, this corresponds to a strain of $h\sim 10^{-27}$ at the detector.
Typical GW signals consist of two major contributions,
a high frequency signal, corresponding to the f-mode around $1$--$2$~kHz and a low frequency contribution associated to Alfv\'en oscillations in the NS core around $f\sim 100$~Hz, which depends on the magnetic field strength. 

Radio pulsars known for their very stable spin periods can occasionally
undergo a  sudden increase in their rotation frequency. These are called glitches
and several hundred glitches have been observed in over 100 pulsars \cite{espinoza11}. There two main physical models for the explanation of glitches and both models involve a substantial rearrangement of the NS
structure on a short time scale, therefore one can expect a bursts of
gravitational radiation, both from the glitches themselves and from
subsequent relaxation of the NS structure. The dynamics
and duration of these phases, however, is to date not well understood
and the predictions of the emission of GWs and their
detectability vary widely. The most pessimistic ones expect that not even 3rd
generation instruments can detect the signal \cite{sidery10}, moderately
optimistically ones \cite{keer15} predict the signals to be detectable by
the ET while the most optimistic ones
\cite{melatos15,bennett10,prix11} expect that
the signals should be marginally detectable even by Advanced
LIGO/Virgo. One can therefore from both  detection, or non-detection by 3rd
generation instruments expect to constrain the physics of the NS interior.

Given the predicted strengths, sensitivity improvements of factors $10$--$100$ compared to current facilities are required for the detection and study of these sources. This necessity extends for all of the sources discussed in this white papers: uncovering the inner workings of stellar core collapse events and supernovae and reliably studying the equation of state of both hot and cold ultra-dense nuclear matter, supported by a sample of several sources and different types of signals, can be achieved only with a dramatic advance of GW sensitivity at high frequencies, that only 3rd-generation ground-based GW detectors can deliver.

\chapterimage{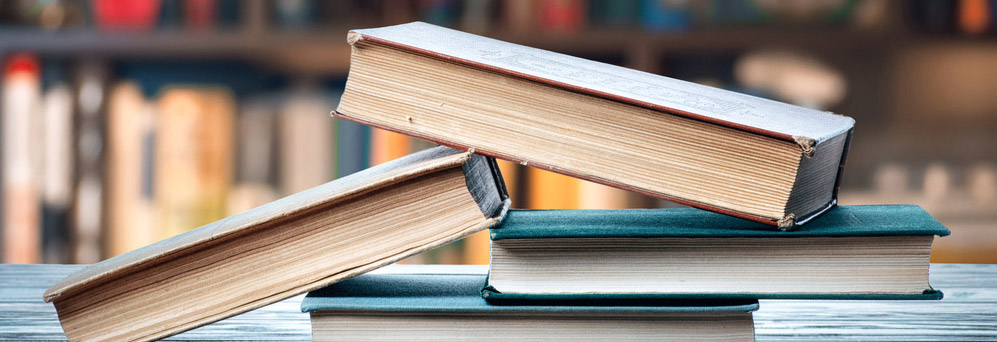} 
\bibliographystyle{utphys}
\bibliography{wp,3g}

\providecommand{\href}[2]{#2}\begingroup\raggedright\begin{thebibliography}{10}

\bibitem{Abbott:2016blz}
{\bfseries LIGO Scientific, Virgo} Collaboration, B.~P. Abbott {\em et~al.},
  ``{Observation of Gravitational Waves from a Binary Black Hole Merger},''
  \href{http://dx.doi.org/10.1103/PhysRevLett.116.061102}{{\em Phys. Rev.
  Lett.} {\bfseries 116} no.~6, (2016) 061102}.

\bibitem{LIGOScientific:2018mvr}
{\bfseries LIGO Scientific, Virgo} Collaboration, B.~P. Abbott {\em et~al.},
  ``{GWTC-1: A Gravitational-Wave Transient Catalog of Compact Binary Mergers
  Observed by LIGO and Virgo during the First and Second Observing Runs},''
\href{http://arxiv.org/abs/1811.12907}{{\ttfamily arXiv:1811.12907
  [astro-ph.HE]}}.

\bibitem{GBM:2017lvd}
{\bfseries LIGO Scientific, Virgo, Fermi GBM, INTEGRAL, IceCube, AstroSat
  Cadmium Zinc Telluride Imager Team, IPN, Insight-Hxmt, ANTARES, Swift, AGILE
  Team, 1M2H Team, Dark Energy Camera GW-EM, DES, DLT40, GRAWITA, Fermi-LAT,
  ATCA, ASKAP, Las Cumbres Observatory Group, OzGrav, DWF (Deeper Wider Faster
  Program), AST3, CAASTRO, VINROUGE, MASTER, J-GEM, GROWTH, JAGWAR,
  CaltechNRAO, TTU-NRAO, NuSTAR, Pan-STARRS, MAXI Team, TZAC Consortium, KU,
  Nordic Optical Telescope, ePESSTO, GROND, Texas Tech University, SALT Group,
  TOROS, BOOTES, MWA, CALET, IKI-GW Follow-up, H.E.S.S., LOFAR, LWA, HAWC,
  Pierre Auger, ALMA, Euro VLBI Team, Pi of Sky, Chandra Team at McGill
  University, DFN, ATLAS Telescopes, High Time Resolution Universe Survey,
  RIMAS, RATIR, SKA South Africa/MeerKAT} Collaboration, B.~P. Abbott {\em
  et~al.}, ``{Multi-messenger Observations of a Binary Neutron Star Merger},''
  \href{http://dx.doi.org/10.3847/2041-8213/aa91c9}{{\em Astrophys. J. Lett.}
  {\bfseries 848} no.~2, (2017) L12}.

\bibitem{murphy:09}
J.~W. {Murphy}, C.~D. {Ott}, and A.~{Burrows}, ``{A Model for Gravitational
  Wave Emission from Neutrino-Driven Core-Collapse Supernovae},''
  \href{http://dx.doi.org/10.1088/0004-637X/707/2/1173}{{\em \apj} {\bfseries
  707} (Dec., 2009) 1173--1190}.

\bibitem{yakunin:15}
K.~N. {Yakunin}, A.~{Mezzacappa}, P.~{Marronetti}, S.~{Yoshida}, S.~W.
  {Bruenn}, W.~R. {Hix}, E.~J. {Lentz}, O.~E. {Bronson Messer}, J.~A. {Harris},
  E.~{Endeve}, J.~M. {Blondin}, and E.~J. {Lingerfelt}, ``{Gravitational wave
  signatures of ab initio two-dimensional core collapse supernova explosion
  models for 12 -25 M$_{?}$ stars},''
  \href{http://dx.doi.org/10.1103/PhysRevD.92.084040}{{\em \prd} {\bfseries 92}
  no.~8, (Oct., 2015) 084040}.

\bibitem{morozova}
V.~{Morozova}, D.~{Radice}, A.~{Burrows}, and D.~{Vartanyan}, ``{The
  Gravitational Wave Signal from Core-collapse Supernovae},''
  \href{http://dx.doi.org/10.3847/1538-4357/aac5f1}{{\em \apj} {\bfseries 861}
  (July, 2018) 10}.

\bibitem{MuJaMa13}
B.~{M{\"u}ller}, H.-T. {Janka}, and A.~{Marek}, ``{A New Multi-dimensional
  General Relativistic Neutrino Hydrodynamics Code of Core-collapse Supernovae.
  III. Gravitational Wave Signals from Supernova Explosion Models},''
  \href{http://dx.doi.org/10.1088/0004-637X/766/1/43}{{\em \apj} {\bfseries
  766} (Mar., 2013) 43}.

\bibitem{kuroda:14}
T.~{Kuroda}, T.~{Takiwaki}, and K.~{Kotake}, ``{Gravitational wave signatures
  from low-mode spiral instabilities in rapidly rotating supernova cores},''
  \href{http://dx.doi.org/10.1103/PhysRevD.89.044011}{{\em \prd} {\bfseries 89}
  no.~4, (Feb., 2014) 044011}.

\bibitem{mueller:17}
B.~{M{\"u}ller}, T.~{Melson}, A.~{Heger}, and H.-T. {Janka}, ``{Supernova
  simulations from a 3D progenitor model - Impact of perturbations and
  evolution of explosion properties},''
  \href{http://dx.doi.org/10.1093/mnras/stx1962}{{\em \mnras} {\bfseries 472}
  (Nov., 2017) 491--513}.

\bibitem{andresen:17}
H.~{Andresen}, B.~{M{\"u}ller}, E.~{M{\"u}ller}, and H.-T. {Janka},
  ``{Gravitational wave signals from 3D neutrino hydrodynamics simulations of
  core-collapse supernovae},''
  \href{http://dx.doi.org/10.1093/mnras/stx618}{{\em \mnras} {\bfseries 468}
  (June, 2017) 2032--2051}.

\bibitem{dimmelmeier:07}
H.~{Dimmelmeier}, C.~D. {Ott}, H.-T. {Janka}, A.~{Marek}, and E.~{M{\"u}ller},
  ``{Generic Gravitational-Wave Signals from the Collapse of Rotating Stellar
  Cores},'' \href{http://dx.doi.org/10.1103/PhysRevLett.98.251101}{{\em
  Physical Review Letters} {\bfseries 98} no.~25, (June, 2007) 251101}.

\bibitem{dimmelmeier:08}
H.~{Dimmelmeier}, C.~D. {Ott}, A.~{Marek}, and H.-T. {Janka}, ``{Gravitational
  wave burst signal from core collapse of rotating stars},''
  \href{http://dx.doi.org/10.1103/PhysRevD.78.064056}{{\em \prd} {\bfseries 78}
  no.~6, (Sept., 2008) 064056}.

\bibitem{richers:17}
S.~{Richers}, C.~D. {Ott}, E.~{Abdikamalov}, E.~{O'Connor}, and C.~{Sullivan},
  ``{Equation of state effects on gravitational waves from rotating core
  collapse},'' \href{http://dx.doi.org/10.1103/PhysRevD.95.063019}{{\em \prd}
  {\bfseries 95} no.~6, (Mar., 2017) 063019}.

\bibitem{ZwMu97}
T.~{Zwerger} and E.~{Mueller}, ``{Dynamics and gravitational wave signature of
  axisymmetric rotational core collapse.},'' {\em \aap} {\bfseries 320} (Apr.,
  1997) 209--227.

\bibitem{abdikamalov:14}
E.~{Abdikamalov}, S.~{Gossan}, A.~M. {DeMaio}, and C.~D. {Ott}, ``{Measuring
  the angular momentum distribution in core-collapse supernova progenitors with
  gravitational waves},''
  \href{http://dx.doi.org/10.1103/PhysRevD.90.044001}{{\em \prd} {\bfseries 90}
  no.~4, (Aug., 2014) 044001}.

\bibitem{fuller:15}
J.~{Fuller}, H.~{Klion}, E.~{Abdikamalov}, and C.~D. {Ott}, ``{Supernova
  seismology: gravitational wave signatures of rapidly rotating core
  collapse},'' \href{http://dx.doi.org/10.1093/mnras/stv698}{{\em \mnras}
  {\bfseries 450} (June, 2015) 414--427}.

\bibitem{kotake:09}
K.~{Kotake}, W.~{Iwakami}, N.~{Ohnishi}, and S.~{Yamada}, ``{Ray-Tracing
  Analysis of Anisotropic Neutrino Radiation for Estimating Gravitational Waves
  in Core-Collapse Supernovae},''
  \href{http://dx.doi.org/10.1088/0004-637X/704/2/951}{{\em \apj} {\bfseries
  704} (Oct., 2009) 951--963}.

\bibitem{mueller:12}
E.~{M{\"u}ller}, H.-T. {Janka}, and A.~{Wongwathanarat}, ``{Parametrized 3D
  models of neutrino-driven supernova explosions. Neutrino emission asymmetries
  and gravitational-wave signals},''
  \href{http://dx.doi.org/10.1051/0004-6361/201117611}{{\em \aap} {\bfseries
  537} (Jan., 2012) A63}.

\bibitem{MuRaBu04}
E.~{M{\"u}ller}, M.~{Rampp}, R.~{Buras}, H.-T. {Janka}, and D.~H. {Shoemaker},
  ``{Toward Gravitational Wave Signals from Realistic Core-Collapse Supernova
  Models},'' \href{http://dx.doi.org/10.1086/381360}{{\em \apj} {\bfseries 603}
  (Mar., 2004) 221--230}.

\bibitem{yakunin:10}
K.~N. {Yakunin}, P.~{Marronetti}, A.~{Mezzacappa}, S.~W. {Bruenn}, C.-T. {Lee},
  M.~A. {Chertkow}, W.~R. {Hix}, J.~M. {Blondin}, E.~J. {Lentz}, O.~E.~B.
  {Messer}, and S.~{Yoshida}, ``{Gravitational waves from core collapse
  supernovae},'' \href{http://dx.doi.org/10.1088/0264-9381/27/19/194005}{{\em
  Classical and Quantum Gravity} {\bfseries 27} no.~19, (Oct., 2010) 194005}.

\bibitem{mueller:13}
B.~{M{\"u}ller}, H.-T. {Janka}, and A.~{Marek}, ``{A New Multi-dimensional
  General Relativistic Neutrino Hydrodynamics Code of Core-collapse Supernovae.
  III. Gravitational Wave Signals from Supernova Explosion Models},''
  \href{http://dx.doi.org/10.1088/0004-637X/766/1/43}{{\em \apj} {\bfseries
  766} (Mar., 2013) 43}.

\bibitem{mueller:97}
E.~{Mueller} and H.-T. {Janka}, ``{Gravitational radiation from convective
  instabilities in Type II supernova explosions.},'' {\em \aap} {\bfseries 317}
  (Jan., 1997) 140--163.

\bibitem{marek_08}
A.~{Marek}, H.~{Janka}, and E.~{M{\"u}ller}, ``{Equation-of-state dependent
  features in shock-oscillation modulated neutrino and gravitational-wave
  signals from supernovae},''
  \href{http://dx.doi.org/10.1051/0004-6361/200810883}{{\em \aap} {\bfseries
  496} (Mar., 2009) 475--494}.

\bibitem{torres:18}
A.~{Torres-Forn{\'e}}, P.~{Cerd{\'a}-Dur{\'a}n}, A.~{Passamonti}, and J.~A.
  {Font}, ``{Towards asteroseismology of core-collapse supernovae with
  gravitational-wave observations - I. Cowling approximation},''
  \href{http://dx.doi.org/10.1093/mnras/stx3067}{{\em \mnras} {\bfseries 474}
  (Mar., 2018) 5272--5286}.

\bibitem{murphy:08}
J.~W. {Murphy} and A.~{Burrows}, ``{Criteria for Core-Collapse Supernova
  Explosions by the Neutrino Mechanism},''
  \href{http://dx.doi.org/10.1086/592214}{{\em \apj} {\bfseries 688} (Dec.,
  2008) 1159--1175}.

\bibitem{cedraduran:13}
P.~{Cerd{\'a}-Dur{\'a}n}, N.~{DeBrye}, M.~A. {Aloy}, J.~A. {Font}, and
  M.~{Obergaulinger}, ``{Gravitational Wave Signatures in Black Hole Forming
  Core Collapse},'' \href{http://dx.doi.org/10.1088/2041-8205/779/2/L18}{{\em
  \apjl} {\bfseries 779} (Dec., 2013) L18}.

\bibitem{BlMe07}
J.~M. {Blondin} and A.~{Mezzacappa}, ``{Pulsar spins from an instability in the
  accretion shock of supernovae},''
  \href{http://dx.doi.org/10.1038/nature05428}{{\em \nat} {\bfseries 445}
  (Jan., 2007) 58--60}.

\bibitem{kuroda:16}
T.~{Kuroda}, K.~{Kotake}, and T.~{Takiwaki}, ``{A New Gravitational-wave
  Signature from Standing Accretion Shock Instability in Supernovae},''
  \href{http://dx.doi.org/10.3847/2041-8205/829/1/L14}{{\em \apjl} {\bfseries
  829} (Sept., 2016) L14}.

\bibitem{kotake:13}
K.~{Kotake}, ``{Multiple physical elements to determine the gravitational-wave
  signatures of core-collapse supernovae},''
  \href{http://dx.doi.org/10.1016/j.crhy.2013.01.008}{{\em Comptes Rendus
  Physique} {\bfseries 14} (Apr., 2013) 318--351}.

\bibitem{abe:2016}
K.~{Abe}, Y.~{Haga}, Y.~{Hayato}, M.~{Ikeda}, K.~{Iyogi}, J.~{Kameda},
  Y.~{Kishimoto}, M.~{Miura}, S.~{Moriyama}, M.~{Nakahata}, Y.~{Nakano},
  S.~{Nakayama}, H.~{Sekiya}, M.~{Shiozawa}, Y.~{Suzuki}, A.~{Takeda},
  H.~{Tanaka}, T.~{Tomura}, K.~{Ueno}, R.~A. {Wendell}, T.~{Yokozawa},
  T.~{Irvine}, T.~{Kajita}, I.~{Kametani}, K.~{Kaneyuki}, K.~P. {Lee},
  T.~{McLachlan}, Y.~{Nishimura}, E.~{Richard}, K.~{Okumura}, L.~{Labarga},
  P.~{Fernandez}, S.~{Berkman}, H.~A. {Tanaka}, S.~{Tobayama}, J.~{Gustafson},
  E.~{Kearns}, J.~L. {Raaf}, J.~L. {Stone}, L.~R. {Sulak}, M.~{Goldhaber},
  G.~{Carminati}, W.~R. {Kropp}, S.~{Mine}, P.~{Weatherly}, A.~{Renshaw}, M.~B.
  {Smy}, H.~W. {Sobel}, V.~{Takhistov}, K.~S. {Ganezer}, B.~L. {Hartfiel},
  J.~{Hill}, W.~E. {Keig}, N.~{Hong}, J.~Y. {Kim}, I.~T. {Lim}, T.~{Akiri},
  A.~{Himmel}, K.~{Scholberg}, C.~W. {Walter}, T.~{Wongjirad}, T.~{Ishizuka},
  S.~{Tasaka}, J.~S. {Jang}, J.~G. {Learned}, S.~{Matsuno}, S.~N. {Smith},
  T.~{Hasegawa}, T.~{Ishida}, T.~{Ishii}, T.~{Kobayashi}, T.~{Nakadaira},
  K.~{Nakamura}, Y.~{Oyama}, K.~{Sakashita}, T.~{Sekiguchi}, T.~{Tsukamoto},
  A.~T. {Suzuki}, Y.~{Takeuchi}, C.~{Bronner}, S.~{Hirota}, K.~{Huang},
  K.~{Ieki}, T.~{Kikawa}, A.~{Minamino}, A.~{Murakami}, T.~{Nakaya},
  K.~{Suzuki}, S.~{Takahashi}, K.~{Tateishi}, Y.~{Fukuda}, K.~{Choi},
  Y.~{Itow}, G.~{Mitsuka}, P.~{Mijakowski}, J.~{Hignight}, J.~{Imber}, C.~K.
  {Jung}, C.~{Yanagisawa}, M.~J. {Wilking}, H.~{Ishino}, A.~{Kibayashi},
  Y.~{Koshio}, T.~{Mori}, M.~{Sakuda}, R.~{Yamaguchi}, T.~{Yano}, Y.~{Kuno},
  R.~{Tacik}, S.~B. {Kim}, H.~{Okazawa}, Y.~{Choi}, K.~{Nishijima},
  M.~{Koshiba}, Y.~{Suda}, Y.~{Totsuka}, M.~{Yokoyama}, K.~{Martens},
  L.~{Marti}, M.~R. {Vagins}, J.~F. {Martin}, P.~{de Perio}, A.~{Konaka},
  S.~{Chen}, Y.~{Zhang}, K.~{Connolly}, and R.~J. {Wilkes}, ``{Real-time
  supernova neutrino burst monitor at Super-Kamiokande},''
  \href{http://dx.doi.org/10.1016/j.astropartphys.2016.04.003}{{\em
  Astroparticle Physics} {\bfseries 81} (Aug., 2016) 39--48}.

\bibitem{ankowski:2016}
A.~{Ankowski}, J.~{Beacom}, O.~{Benhar}, S.~{Chen}, J.~{Cherry}, Y.~{Cui},
  A.~{Friedland}, I.~{Gil-Botella}, A.~{Haghighat}, S.~{Horiuchi}, P.~{Huber},
  J.~{Kneller}, R.~{Laha}, S.~{Li}, J.~{Link}, A.~{Lovato}, O.~{Macias},
  C.~{Mariani}, A.~{Mezzacappa}, E.~{O'Connor}, E.~{O'Sullivan}, A.~{Rubbia},
  K.~{Scholberg}, and T.~{Takeuchi}, ``{Supernova Physics at DUNE},'' {\em
  ArXiv e-prints} (Aug., 2016) ,
  \href{http://arxiv.org/abs/1608.07853}{{\ttfamily arXiv:1608.07853
  [hep-ex]}}.

\bibitem{lu:2015}
J.-S. {Lu}, J.~{Cao}, Y.-F. {Li}, and S.~{Zhou}, ``{Constraining absolute
  neutrino masses via detection of galactic supernova neutrinos at JUNO},''
  \href{http://dx.doi.org/10.1088/1475-7516/2015/05/044}{{\em \jcap} {\bfseries
  5} (May, 2015) 044}.

\bibitem{abbasi:2011}
R.~{Abbasi}, Y.~{Abdou}, T.~{Abu-Zayyad}, M.~{Ackermann}, J.~{Adams}, J.~A.
  {Aguilar}, M.~{Ahlers}, M.~M. {Allen}, D.~{Altmann}, K.~{Andeen}, and et~al.,
  ``{IceCube sensitivity for low-energy neutrinos from nearby supernovae},''
  \href{http://dx.doi.org/10.1051/0004-6361/201117810}{{\em \aap} {\bfseries
  535} (Nov., 2011) A109}.

\bibitem{Agafonova:2007hn}
N.~{\relax Yu}. Agafonova {\em et~al.}, ``{On-line recognition of supernova
  neutrino bursts in the LVD detector},''
  \href{http://dx.doi.org/10.1016/j.astropartphys.2007.09.005}{{\em Astropart.
  Phys.} {\bfseries 28} (2008) 516--522}.

\bibitem{Cadonati:2000kq}
L.~Cadonati, F.~P. Calaprice, and M.~C. Chen, ``{Supernova neutrino detection
  in borexino},'' \href{http://dx.doi.org/10.1016/S0927-6505(01)00129-3}{{\em
  Astropart. Phys.} {\bfseries 16} (2002) 361--372}.

\bibitem{Tolich:2011zz}
K.~Tolich, ``{Supernova detection with KamLAND},''
  \href{http://dx.doi.org/10.1016/j.nuclphysbps.2011.10.005}{{\em Nucl. Phys.
  Proc. Suppl.} {\bfseries 221} (2011) 355}.

\bibitem{kuroda:17}
T.~{Kuroda}, K.~{Kotake}, K.~{Hayama}, and T.~{Takiwaki}, ``{Correlated
  Signatures of Gravitational-wave and Neutrino Emission in Three-dimensional
  General-relativistic Core-collapse Supernova Simulations},''
  \href{http://dx.doi.org/10.3847/1538-4357/aa988d}{{\em \apj} {\bfseries 851}
  (Dec., 2017) 62}.

\bibitem{tk18}
T.~{Takiwaki} and K.~{Kotake}, ``{Anisotropic emission of neutrino and
  gravitational-wave signals from rapidly rotating core-collapse supernovae},''
  \href{http://dx.doi.org/10.1093/mnrasl/sly008}{{\em \mnras} {\bfseries 475}
  (Mar., 2018) L91--L95}.

\bibitem{MM}
K.~{Nakamura}, S.~{Horiuchi}, M.~{Tanaka}, K.~{Hayama}, T.~{Takiwaki}, and
  K.~{Kotake}, ``{Multimessenger signals of long-term core-collapse supernova
  simulations: synergetic observation strategies},''
  \href{http://dx.doi.org/10.1093/mnras/stw1453}{{\em \mnras} {\bfseries 461}
  (Sept., 2016) 3296--3313}.

\bibitem{ott:12}
C.~D. {Ott}, E.~{Abdikamalov}, E.~{O'Connor}, C.~{Reisswig}, R.~{Haas},
  P.~{Kalmus}, S.~{Drasco}, A.~{Burrows}, and E.~{Schnetter}, ``{Correlated
  gravitational wave and neutrino signals from general-relativistic rapidly
  rotating iron core collapse},''
  \href{http://dx.doi.org/10.1103/PhysRevD.86.024026}{{\em \prd} {\bfseries 86}
  no.~2, (July, 2012) 024026}.

\bibitem{yokozawa:15}
T.~{Yokozawa}, M.~{Asano}, T.~{Kayano}, Y.~{Suwa}, N.~{Kanda}, Y.~{Koshio}, and
  M.~R. {Vagins}, ``{Probing the Rotation of Core-collapse Supernova with a
  Concurrent Analysis of Gravitational Waves and Neutrinos},''
  \href{http://dx.doi.org/10.1088/0004-637X/811/2/86}{{\em \apj} {\bfseries
  811} (Oct., 2015) 86}.

\bibitem{ott_11}
C.~D. {Ott}, C.~{Reisswig}, E.~{Schnetter}, E.~{O'Connor}, U.~{Sperhake},
  F.~{L{\"o}ffler}, P.~{Diener}, E.~{Abdikamalov}, I.~{Hawke}, and
  A.~{Burrows}, ``{Dynamics and Gravitational Wave Signature of Collapsar
  Formation},'' \href{http://dx.doi.org/10.1103/PhysRevLett.106.161103}{{\em
  Physical Review Letters} {\bfseries 106} no.~16, (Apr., 2011) 161103--+}.

\bibitem{cerda_13}
P.~{Cerd{\'a}-Dur{\'a}n}, N.~{DeBrye}, M.~A. {Aloy}, J.~A. {Font}, and
  M.~{Obergaulinger}, ``{Gravitational Wave Signatures in Black Hole Forming
  Core Collapse},'' \href{http://dx.doi.org/10.1088/2041-8205/779/2/L18}{{\em
  \apjl} {\bfseries 779} (Dec., 2013) L18}.

\bibitem{pan:17}
K.-C. {Pan}, M.~{Liebend{\"o}rfer}, S.~M. {Couch}, and F.-K. {Thielemann},
  ``{Equation of State Dependent Dynamics and Multi-messenger Signals from
  Stellar-mass Black Hole Formation},''
  \href{http://dx.doi.org/10.3847/1538-4357/aab71d}{{\em \apj} {\bfseries 857}
  (Apr., 2018) 13}.

\bibitem{yakunin:17}
K.~N. {Yakunin}, A.~{Mezzacappa}, P.~{Marronetti}, E.~J. {Lentz}, S.~W.
  {Bruenn}, W.~R. {Hix}, O.~E.~B. {Messer}, E.~{Endeve}, J.~M. {Blondin}, and
  J.~A. {Harris}, ``{The Gravitational Wave Signal of a Core Collapse Supernova
  Explosion of a 15M$\_\odot$ Star},'' {\em ArXiv e-prints} (Jan., 2017) .

\bibitem{Hild:2010id}
S.~Hild {\em et~al.}, ``{Sensitivity Studies for Third-Generation Gravitational
  Wave Observatories},''
  \href{http://dx.doi.org/10.1088/0264-9381/28/9/094013}{{\em Class. Quant.
  Grav.} {\bfseries 28} (2011) 094013}.

\bibitem{Dwyer:2014fpa}
S.~Dwyer, D.~Sigg, S.~W. Ballmer, L.~Barsotti, N.~Mavalvala, and M.~Evans,
  ``{Gravitational wave detector with cosmological reach},''
  \href{http://dx.doi.org/10.1103/PhysRevD.91.082001}{{\em Phys. Rev.}
  {\bfseries D91} no.~8, (2015) 082001}.

\bibitem{Aasi:2013wya}
{\bfseries KAGRA, LIGO Scientific, Virgo} Collaboration, B.~P. Abbott {\em
  et~al.}, ``{Prospects for Observing and Localizing Gravitational-Wave
  Transients with Advanced LIGO, Advanced Virgo and KAGRA},''
  \href{http://dx.doi.org/10.1007/s41114-018-0012-9}{{\em Living Rev. Rel.}
  {\bfseries 21} no.~1, (2018) 3}.

\bibitem{Gossan:2015xda}
S.~E. Gossan, P.~Sutton, A.~Stuver, M.~Zanolin, K.~Gill, and C.~D. Ott,
  ``{Observing Gravitational Waves from Core-Collapse Supernovae in the
  Advanced Detector Era},''
  \href{http://dx.doi.org/10.1103/PhysRevD.93.042002}{{\em Phys. Rev. D}
  {\bfseries 93} no.~4, (2016) 042002}.

\bibitem{Heng:2008ww}
I.~S. Heng, ``{Rotating stellar core-collapse waveform decomposition: A
  principal component analysis approach},''
  \href{http://dx.doi.org/10.1088/0264-9381/26/10/105005}{{\em Class. Quant.
  Grav.} {\bfseries 26} (2009) 105005}.

\bibitem{powell:16}
J.~{Powell}, S.~E. {Gossan}, J.~{Logue}, and I.~S. {Heng}, ``{Inferring the
  core-collapse supernova explosion mechanism with gravitational waves},''
  \href{http://dx.doi.org/10.1103/PhysRevD.94.123012}{{\em \prd} {\bfseries 94}
  no.~12, (Dec., 2016) 123012}.

\bibitem{2017MPLA...3230035R}
K.~{Riles}, ``{Recent searches for continuous gravitational waves},''
  \href{http://dx.doi.org/10.1142/S021773231730035X}{{\em Modern Physics
  Letters A} {\bfseries 32} (Dec., 2017) 1730035--685}.

\bibitem{1998ApJ...502..708A}
N.~{Andersson}, ``{A New Class of Unstable Modes of Rotating Relativistic
  Stars},'' \href{http://dx.doi.org/10.1086/305919}{{\em \apj} {\bfseries 502}
  (Aug., 1998) 708--713}.

\bibitem{1998ApJ...502..714F}
J.~L. {Friedman} and S.~M. {Morsink}, ``{Axial Instability of Rotating
  Relativistic Stars},'' \href{http://dx.doi.org/10.1086/305920}{{\em \apj}
  {\bfseries 502} (Aug., 1998) 714--720}.

\bibitem{Pisarski:2019vxw}
{\bfseries LIGO Scientific, Virgo} Collaboration, B.~P. Abbott {\em et~al.},
  ``{All-sky Search for Continuous Gravitational Waves from Isolated Neutron
  Stars using Advanced LIGO O2 Data},''
\href{http://arxiv.org/abs/1903.01901}{{\ttfamily arXiv:1903.01901
  [astro-ph.HE]}}.

\bibitem{Authors:2019ztc}
{\bfseries LIGO Scientific, Virgo} Collaboration, ``{Searches for Gravitational
  Waves from Known Pulsars at Two Harmonics in 2015-2017 LIGO Data},''
\href{http://arxiv.org/abs/1902.08507}{{\ttfamily arXiv:1902.08507
  [astro-ph.HE]}}.

\bibitem{Ruderman:1969}
M.~Ruderman, ``Neutron starquakes and pulsar periods,'' {\em Nature (London)}
  {\bfseries 223} no.~5206, (1969) 597.

\bibitem{JohnsonMcDaniel:2012wg}
N.~K. Johnson-McDaniel and B.~J. Owen, ``{Maximum elastic deformations of
  relativistic stars},''
  \href{http://dx.doi.org/10.1103/PhysRevD.88.044004}{{\em Phys. Rev. D}
  {\bfseries 88} (2013) 044004}.

\bibitem{Ushomirsky:2000ax}
G.~Ushomirsky, C.~Cutler, and L.~Bildsten, ``{Deformations of accreting neutron
  star crusts and gravitational wave emission},''
  \href{http://dx.doi.org/10.1046/j.1365-8711.2000.03938.x}{{\em Mon. Not. R.
  Astron. Soc.} {\bfseries 319} (2000) 902}.

\bibitem{Douchin:2001sv}
F.~Douchin and P.~Haensel, ``{A unified equation of state of dense matter and
  neutron star structure},''
  \href{http://dx.doi.org/10.1051/0004-6361:20011402}{{\em Astron. Astrophys.}
  {\bfseries 380} (2001) 151}.

\bibitem{Abbott:2018exr}
{\bfseries LIGO Scientific, Virgo} Collaboration, B.~P. Abbott {\em et~al.},
  ``{GW170817: Measurements of neutron star radii and equation of state},''
  \href{http://dx.doi.org/10.1103/PhysRevLett.121.161101}{{\em Phys. Rev.
  Lett.} {\bfseries 121} no.~16, (2018) 161101}.

\bibitem{Woan:2018tey}
G.~Woan, M.~D. Pitkin, B.~Haskell, D.~I. Jones, and P.~D. Lasky, ``{Evidence
  for a Minimum Ellipticity in Millisecond Pulsars},''
  \href{http://dx.doi.org/10.3847/2041-8213/aad86a}{{\em Astrophys. J. Lett.}
  {\bfseries 863} no.~2, (2018) L40}.

\bibitem{2005ApJ...623.1044M}
A.~{Melatos} and D.~J.~B. {Payne}, ``{Gravitational Radiation from an Accreting
  Millisecond Pulsar with a Magnetically Confined Mountain},''
  \href{http://dx.doi.org/10.1086/428600}{{\em \apj} {\bfseries 623} (Apr.,
  2005) 1044--1050}.

\bibitem{1998ApJ...501L..89B}
L.~{Bildsten}, ``{Gravitational Radiation and Rotation of Accreting Neutron
  Stars},'' \href{http://dx.doi.org/10.1086/311440}{{\em \apjl} {\bfseries 501}
  (July, 1998) L89--L93}.

\bibitem{2014MNRAS.442.1786A}
N.~{Andersson}, D.~I. {Jones}, and W.~C.~G. {Ho}, ``{Implications of an r mode
  in XTE J1751-305: mass, radius and spin evolution},''
  \href{http://dx.doi.org/10.1093/mnras/stu870}{{\em \mnras} {\bfseries 442}
  (Aug., 2014) 1786--1793}.

\bibitem{2014MNRAS.442.3037L}
U.~{Lee}, ``{Excitation of a non-radial mode in a millisecond X-ray pulsar XTE
  J1751-305},'' \href{http://dx.doi.org/10.1093/mnras/stu1077}{{\em \mnras}
  {\bfseries 442} (Aug., 2014) 3037--3043}.

\bibitem{2014ApJ...784...72S}
T.~{Strohmayer} and S.~{Mahmoodifar}, ``{A Non-radial Oscillation Mode in an
  Accreting Millisecond Pulsar?},''
  \href{http://dx.doi.org/10.1088/0004-637X/784/1/72}{{\em \apj} {\bfseries
  784} (Mar., 2014) 72}.

\bibitem{2014ApJ...793L..38S}
T.~{Strohmayer} and S.~{Mahmoodifar}, ``{Discovery of a Neutron Star
  Oscillation Mode During a Superburst},''
  \href{http://dx.doi.org/10.1088/2041-8205/793/2/L38}{{\em \apjl} {\bfseries
  793} (Oct., 2014) L38}.

\bibitem{2003Natur.424...42C}
D.~{Chakrabarty}, E.~H. {Morgan}, M.~P. {Muno}, D.~K. {Galloway},
  R.~{Wijnands}, M.~{van der Klis}, and C.~B. {Markwardt}, ``{Nuclear-powered
  millisecond pulsars and the maximum spin frequency of neutron stars},''
  \href{http://dx.doi.org/10.1038/nature01732}{{\em \nat} {\bfseries 424}
  (July, 2003) 42--44}.

\bibitem{2008AIPC.1068...67C}
D.~{Chakrabarty}, \href{http://dx.doi.org/10.1063/1.3031208}{``{The spin
  distribution of millisecond X-ray pulsars},''} in {\em American Institute of
  Physics Conference Series}, R.~{Wijnands}, D.~{Altamirano}, P.~{Soleri},
  N.~{Degenaar}, N.~{Rea}, P.~{Casella}, A.~{Patruno}, and M.~{Linares}, eds.,
  vol.~1068 of {\em American Institute of Physics Conference Series},
  pp.~67--74.
\newblock Oct., 2008.

\bibitem{2017ApJ...850..106P}
A.~{Patruno}, B.~{Haskell}, and N.~{Andersson}, ``{The Spin Distribution of
  Fast-spinning Neutron Stars in Low-mass X-Ray Binaries: Evidence for Two
  Subpopulations},'' \href{http://dx.doi.org/10.3847/1538-4357/aa927a}{{\em
  \apj} {\bfseries 850} (Nov., 2017) 106}.

\bibitem{kaspi17}
V.~M. {Kaspi} and A.~M. {Beloborodov}, ``{Magnetars},''
  \href{http://dx.doi.org/10.1146/annurev-astro-081915-023329}{{\em \araa}
  {\bfseries 55} (Aug., 2017) 261--301}.

\bibitem{turolla15}
R.~{Turolla}, S.~{Zane}, and A.~L. {Watts}, ``{Magnetars: the physics behind
  observations. A review},''
  \href{http://dx.doi.org/10.1088/0034-4885/78/11/116901}{{\em Reports on
  Progress in Physics} {\bfseries 78} no.~11, (Nov., 2015) 116901}.

\bibitem{lasky15}
P.~D. {Lasky}, A.~{Melatos}, V.~{Ravi}, and G.~{Hobbs}, ``{Pulsar timing noise
  and the minimum observation time to detect gravitational waves with pulsar
  timing arrays},'' \href{http://dx.doi.org/10.1093/mnras/stv540}{{\em \mnras}
  {\bfseries 449} (May, 2015) 3293--3300}.

\bibitem{glampedakis17}
K.~{Glampedakis} and L.~{Gualtieri}, ``{Gravitational Waves from Single Neutron
  Stars: An Advanced Detector Era Survey},''
  \href{http://arxiv.org/abs/1709.07049}{{\ttfamily arXiv:1709.07049
  [astro-ph.HE]}}.

\bibitem{israel05}
G.~L. {Israel}, T.~{Belloni}, L.~{Stella}, Y.~{Rephaeli}, D.~E. {Gruber},
  P.~{Casella}, S.~{Dall'Osso}, N.~{Rea}, M.~{Persic}, and R.~E. {Rothschild},
  ``{The Discovery of Rapid X-Ray Oscillations in the Tail of the SGR 1806-20
  Hyperflare},'' \href{http://dx.doi.org/10.1086/432615}{{\em \apjl} {\bfseries
  628} (July, 2005) L53--L56}.

\bibitem{strohmayer05}
T.~E. {Strohmayer} and A.~L. {Watts}, ``{Discovery of Fast X-Ray Oscillations
  during the 1998 Giant Flare from SGR 1900+14},''
  \href{http://dx.doi.org/10.1086/497911}{{\em \apjl} {\bfseries 632} (Oct.,
  2005) L111--L114}.

\bibitem{strohmayer06}
T.~E. {Strohmayer} and A.~L. {Watts}, ``{The 2004 Hyperflare from SGR 1806-20:
  Further Evidence for Global Torsional Vibrations},''
  \href{http://dx.doi.org/10.1086/508703}{{\em \apj} {\bfseries 653} (Dec.,
  2006) 593--601}.

\bibitem{huppenkothen14a}
D.~{Huppenkothen}, L.~M. {Heil}, A.~L. {Watts}, and E.~{G{\"o}{\u g}{\"u}{\c
  s}}, ``{Quasi-periodic Oscillations in Short Recurring Bursts of Magnetars
  SGR 1806-20 and SGR 1900+14 Observed with RXTE},''
  \href{http://dx.doi.org/10.1088/0004-637X/795/2/114}{{\em \apj} {\bfseries
  795} (Nov., 2014) 114}.

\bibitem{huppenkothen14b}
D.~{Huppenkothen}, C.~{D'Angelo}, A.~L. {Watts}, L.~{Heil}, M.~{van der Klis},
  A.~J. {van der Horst}, C.~{Kouveliotou}, M.~G. {Baring}, E.~{G{\"o}{\u
  g}{\"u}{\c s}}, J.~{Granot}, Y.~{Kaneko}, L.~{Lin}, A.~{von Kienlin}, and
  G.~{Younes}, ``{Quasi-periodic Oscillations in Short Recurring Bursts of the
  Soft Gamma Repeater J1550-5418},''
  \href{http://dx.doi.org/10.1088/0004-637X/787/2/128}{{\em \apj} {\bfseries
  787} (June, 2014) 128}.

\bibitem{espinoza11}
C.~M. {Espinoza}, A.~G. {Lyne}, B.~W. {Stappers}, and M.~{Kramer}, ``{A study
  of 315 glitches in the rotation of 102 pulsars},''
  \href{http://dx.doi.org/10.1111/j.1365-2966.2011.18503.x}{{\em MNRAS}
  {\bfseries 414} (June, 2011) 1679--1704}.

\bibitem{sidery10}
T.~{Sidery}, A.~{Passamonti}, and N.~{Andersson}, ``{The dynamics of pulsar
  glitches: contrasting phenomenology with numerical evolutions},''
  \href{http://dx.doi.org/10.1111/j.1365-2966.2010.16497.x}{{\em MNRAS}
  {\bfseries 405} (June, 2010) 1061--1074}.

\bibitem{keer15}
L.~{Keer} and D.~I. {Jones}, ``{Developing a model for neutron star
  oscillations following starquakes},''
  \href{http://dx.doi.org/10.1093/mnras/stu2123}{{\em MNRAS} {\bfseries 446}
  (Jan., 2015) 865--891}.

\bibitem{melatos15}
A.~{Melatos}, J.~A. {Douglass}, and T.~P. {Simula}, ``{Persistent Gravitational
  Radiation from Glitching Pulsars},''
  \href{http://dx.doi.org/10.1088/0004-637X/807/2/132}{{\em ApJ} {\bfseries
  807} (July, 2015) 132}.

\bibitem{bennett10}
M.~F. {Bennett}, C.~A. {van Eysden}, and A.~{Melatos}, ``{Continuous-wave
  gravitational radiation from pulsar glitch recovery},''
  \href{http://dx.doi.org/10.1111/j.1365-2966.2010.17416.x}{{\em MNRAS}
  {\bfseries 409} (Dec., 2010) 1705--1718}.

\bibitem{prix11}
R.~{Prix}, S.~{Giampanis}, and C.~{Messenger}, ``{Search method for
  long-duration gravitational-wave transients from neutron stars},''
  \href{http://dx.doi.org/10.1103/PhysRevD.84.023007}{{\em Phys. Rev. D}
  {\bfseries 84} no.~2, (July, 2011) 023007}.

\end{thebibliography}\endgroup

\end{document}